\documentstyle[12pt]{article}
\newcommand{\be}{\begin{equation}}
\newcommand{\ee}{\end{equation}}
\date{}

\newcommand{\GeV}{\mbox{GeV}}
\def \lta {\mathrel{\vcenter
     {\hbox{$<$}\nointerlineskip\hbox{$\sim$}}}}

\begin{document}
\begin{titlepage}
\begin{flushright}
HD--THEP--93--30\\
\end{flushright}
\vspace{1.8cm}
\begin{center}
{\bf\LARGE FLOW EQUATIONS FOR N POINT}\\
\vspace{.3cm}
{\bf\LARGE FUNCTIONS}\\
\vspace{.3cm}
{\bf\LARGE AND BOUND STATES}\\
\vspace{1cm}
Ulrich Ellwanger\footnote{Supported by  a DFG Heisenberg fellowship,
e-mail: I96 at VM.URZ.UNI-HEIDELBERG.DE}\\
\vspace{.5cm}
Institut  f\"ur Theoretische Physik\\
Universit\"at Heidelberg\\
Philosophenweg 16, D-69120 Heidelberg, FRG\\
\vspace{3cm}
{\bf Abstract:}\\
\parbox[t]{\textwidth}{We discuss the exact renormalization group or
flow equation for the effective action and its decomposition into one
particle irreducible N point functions. With the help of a truncated
flow equation for the four point function we study the bound state
problem for scalar fields. A combination of analytic and numerical
methods is proposed, which is applied to the Wick-Cutkosky model and a
QCD-motivated interaction. We present results for the bound state
masses and the Bethe-Salpeter wave function. }

\end{center}\end{titlepage}
\newpage

\section{Introduction}
For many considerations in quantum field theory it is a helpful idea to
integrate out high frequency or short distance modes, and to study a
scale-dependent effective theory for the remaining low frequency modes.
This concept was first made concrete on the lattice within the
framework of block spin transformations \cite{1}. When applied to
continuum quantum field theory, it can be cast into the form of exact
renormalization group equations \cite{2}-\cite{10}, which describe the
cutoff dependence of the effective theory in a compact way. Polchinski
and others \cite{2}-\cite{7} employed these ideas in order to study
the dependence of physical Green functions on the UV cutoff with the
aim to simplify perturbative proofs of renormalizability. The
integration of the corresponding flow equations with respect to an
infrared cutoff, on the other hand,  provides a new exact method to
compute effective low energy Lagrangians \cite{8}-\cite{10} or average
actions \cite{11}. Within perturbative or 1/N expansions this scheme
has been applied successfully to the computation of effective
potentials and ``running'' wave function normalizations and allows,
e.g., to investigate cricital phenomena in two and three dimensions
\cite{11}, \cite{12}, high temperature phase transitions \cite{13} and
universality within the Higgs top system \cite{10}.

These results employed, to a large extent, an expansion of the effective
action in powers of momenta (as it is appropriate for investigations of
the effective potential). The first purpose of the present paper is the
derivation of an expansion of the exact flow equations for the effective
action in powers of fields, keeping the momentum dependence exact. This
way one finds an infinite set of coupled differential equation for
one-particle irreducible N point functions.

Then we concentrate on the four point function, whose singularities as
a function of the c.m. energy squared  $s$ encode informations on
possible bound states of theory. We study a truncation of the infinite
set of flow equations, which corresponds to
the  ladder approximation. It leads to a simple flow equation including
just the four point function itself. Already in this approximation we
see, that the flow equation allows to find both the masses and the wave
functions of bound states in a theory,  and the equivalence to the
Bethe-Salpeter equation can be shown.

Next we discuss, as an essential step towards numerical integrations of
the flow equations, the Laplace transform of the four point function
with respect to the Lorentz invariant products of momentum variables.
Numerical methods can then be introduced after  discretization  in this
``Laplace space''. Integration of the flow equation becomes  a simple
algorithm to update the corresponding ``Laplace lattice''
consecutively. The informations on bound states can be obtained from
the numerical results in combination with analytic methods.

As a first application we study the Wick-Cutkosky model, which contains
two massive complex scalars interacting through the exchange of a
massless real scalar and ressembles, in the nonrelativistic limit, to
the Positronium problem. We find good agreement between our results,
both on the coupling constant dependence of the bound state mass and on
the bound state wave function, and known formulas. Our method does not
make use, however, of the $O(4)$ symmetry of the  Wick-Cutkosky model
and can straightforwardly be applied to the  bound state problem in the
case of arbitrary interactions. We present formulas, which allow to
study interactions which correspond to potentials, in the
non-relativistic limit, of the form $r^\alpha$ with $\alpha$ arbitrary.
We end up with results for a QCD-motivated potential  including a
linearly rising confining part.

\section{Flow Equations for N Point Functions}
We will present the general features of the flow equations in the
context of a single scalar field $\varphi$ for simplicity, and we work
in Euclidean space. The aim of the flow equations is the computation of
Green functions within a theory, which is regularized in the UV in
terms of a cutoff $\Lambda$ in the propagator. In addition an infrared
cutoff $k$ is introduced below, and an important role is played by the
corresponding propagator \be\label{2.1} P_k^\Lambda(q^2)\equiv
(R_k^\Lambda(q^2))^{-1}=\frac
{h_\Lambda(q^2)-h_k(q^2)}{q^2+m^2}\ee
with
\begin{eqnarray}\label{2.2}
h_k(q^2)\to 1\quad{\rm for}\quad q^2\ll k^2\nonumber\\
h_k(q^2)\to 0\quad{\rm for}\quad q^2\gg k^2.\end{eqnarray}

The starting point is the UV-regularized  generating functional
of connected Green functions $G^\Lambda(J)$, which can be represented
as
\be\label{2.3}
e^{-G^\Lambda(J)}={\cal N}\int {\cal D}\phi
e^{-\frac{1}{2}(\phi,R^\Lambda
_0\phi)-S^\Lambda_{int}(\phi)+(J,\phi)}.\ee
Here $(J,\phi)$ etc. is a
short-hand notation for  \be\label{2.4}
(J,\phi)\equiv \int\frac{d^4q}{(2\pi)^4}J(q)\phi(-q)\ee
and we have represented the kinetic term in terms of the inverse
UV-regularized propagator $R_0^\Lambda(q^2)$. An alternative and
useful representation of $G^\Lambda(J)$ is given by
\be\label{2.5}
e^{-G^\Lambda(J)}=e^{\frac{1}{2}(J,P^\Lambda_0J)}e^{D^\Lambda_0}e^{-S
^\Lambda_{int}(\phi)}\Bigr\vert_{\phi=P^\Lambda_0\cdot J}\ee
with
\be\label{2.6}
D_0^\Lambda=\frac{1}{2}(P^\Lambda_0\frac{\delta}{\delta\phi},
\frac{\delta}
{\delta\phi}).\ee

Next we introduce an infrared cutoff $k$ and define an effective
interaction $S_{int}(\phi,k)$ by
\be\label{2.7}
e^{-S_{int}(\phi,k)}=e^{D_k^\Lambda}e^{-S^\Lambda_{int}(\phi)},\ee
where $D_k^\Lambda$ is given as in (\ref{2.6}) with $P_0^\Lambda$
replace by $P^\Lambda_k$. Clearly $S_{int}(\phi,k)$ is equal to the
bare interaction $S^\Lambda_{int}(\phi)$ at $k=\Lambda$, and from (\ref
{2.5}) we have
\be\label{2.8}
S_{int}(\phi,0)\Bigr\vert_{\phi=P_0^\Lambda\cdot J}=G^\Lambda(J)+
\frac{1}{2}(J,P_0^\Lambda J).\ee
The flow equation for $S_{int}(\phi,k)$ follows easily after
differentiation of the defining equation (\ref{2.7}) with respect to
$k$ (subsequently we us $\partial_k\equiv d/dk^2)$  \cite{8}, \cite{10}:
\be
\label{2.9}
\partial_kS_{int}(\phi,k)=\frac{1}{2}\int\frac{d^4q}{(2\pi)^4}
\partial_kP_k^\Lambda
(q^2)\cdot\Bigl\lbrace\frac{\delta^2S_{int}(\phi,k)}{\delta\phi(q)
\delta\phi(-q)}
-\frac{\delta S_{int}(\phi,k)}{\delta\phi(q)}\frac{\delta
S_{int}(\phi,k)}{\delta\phi(-q)} \Bigr\rbrace.
\ee

The generating functional of connected Green functions including the
infrared cutoff $k,G_k^\Lambda
(J)$, is related to $S_{int}(\phi,k)$ through
\be\label{2.10}
S_{int}(\phi,k)\Bigr\vert_{\phi=P_k^\Lambda\cdot J}=G_k^\Lambda(J)
+\frac{1}{2}(J,P^\Lambda_kJ).\ee

Inserting (\ref{2.10}) into (\ref{2.9}) and taking into account that
$J$ is implicitly $k$-dependent because of the $\phi=P_k^\Lambda\cdot
J$ prescription in (\ref{2.10}), one finds that $G_k^\Lambda(J)$
satisfies a similar flow equation:
\be \label{2.11}
\partial_kG_k^\Lambda(J)=-\frac{1}{2}\int\frac{d^kq}{(2\pi)^4}
\partial_kR_k^\Lambda
\cdot\Bigl\lbrace\frac{\delta^2G_k^\Lambda(J)}{\delta J(q)\delta J(-q)}
-\frac{\delta G_k^\Lambda(J)}{\delta J(q)}\frac{\delta  G_k^\Lambda(J)}
{\delta J(-q)}\Bigr\rbrace.\ee
In (\ref{2.11}) we have neglected a
$J$-independent term in $G_k^\Lambda(J)$. For many considerations in
quantum field theory, however, it is more convenient to deal with the
generating functional of one-particle irreducible Green functions
$\Gamma(\varphi)$. Its constant part gives the effective potential, and
also all other informations on a theory are encoded in
$\Gamma(\varphi)$ in a more compact way. In the presence of a UV cutoff
$\Lambda$ and an infrared cutoff $k$ the effective action
$\Gamma_k^\Lambda (\varphi)$ is given by the Legendre transform of
$G_k^\Lambda(J)$,
\be\label{2.12}
\Gamma_k^\Lambda(\varphi)=G_k^\Lambda(J)+(J,\varphi).\ee

Inserting the Legendre transformation into (\ref{2.11}) it is possible
to derive the flow equation for $\Gamma_k^\Lambda(\varphi)$. Thereby
the implicit $k$-dependences of $J$ and $\varphi$ have to be taken into
account, but at the end they cancel and one is left with the simple
result
\be\label{2.13}
\partial_k\Gamma^\Lambda_k(\varphi)=\frac{1}{2}\int\frac{d^4q}{(2\pi)^4}
\partial_k R_k^\Lambda(q^2)\left\lbrace\varphi(q)\varphi(-q)
+\left(\frac{\delta^2\Gamma_k^\Lambda(\varphi)}
{\delta\varphi(q)\delta\varphi(-q)}\right)
^{-1}\right\rbrace.\ee
Such an equation has also been found be Wetterich \cite{9} and, in a
different context (as a  differential equation with respect to the UV
cutoff $\Lambda$ instead of $k$) in \cite{7}. If one splits off a bare
kinetic part of $\Gamma_k^\Lambda(\varphi)$,
\be\label{2.14}
\Gamma^\Lambda_k(\varphi)=\frac{1}{2}(\varphi,R_k^\Lambda\varphi)+
\tilde\Gamma_k^\Lambda (\varphi),\ee
one obtains a flow equation for $\tilde\Gamma_k^\Lambda(\varphi)$ of
the form
\be\label{2.15}
\partial_k\tilde\Gamma_k(\varphi)=\frac{1}{2}\int\frac{d^4q}{(2\pi)^4}
\partial_k R_k^\Lambda(q^2)\left(R_k^\Lambda(q^2)+\frac{\delta^2\tilde
\Gamma_k^\Lambda(\varphi)}{\delta \varphi(q)\delta\varphi(-q)}\right)^
{-1}.\ee
In general, however, $\tilde\Gamma_k^\Lambda$ will still contain terms
quadratic in $\varphi$. The boundary condition for $k\to\Lambda$ is most
easily given for $\tilde\Gamma_k(\varphi)$: From the relation between
$\Gamma_k^\Lambda(\varphi)$ and $S_{int}(\phi,k)$ via eqs. (\ref{2.12})
and (\ref{2.10}), and a careful consideration of the limit
$P_k^\Lambda\to0$ for $k\to\Lambda$, one finds
\be\label{2.16}
\tilde\Gamma_\Lambda^\Lambda
(\varphi)=S_{int}^\Lambda(\phi)\Bigr|_{\phi=\varphi}\ee
where $S_{int}^\Lambda(\phi)$ is the bare action of eqs. (\ref{2.3}{)
or (\ref{2.5}).

Let us now consider an expansion of $\Gamma_k^\Lambda$ in powers of
$\varphi$, with one particle irreducible $N$ point functions $\Gamma_N$
as coefficients.

For simplicity we will assume a discrete symmetry $\varphi\to-\varphi$
such that only even powers of $\varphi$ appear. Instead of writing the
convolutions over momenta, we switch to an tensor notation (as
appropriate in the case of discretized momenta as, e.g., on a torus)
according to
\be\label{2.17} \varphi(q)\to\varphi_i,\
\Gamma_N(q_1...q_N)\to\Gamma_N^{i_1...i_N},\int
\frac{d^4q}{(2\pi)^"}\to\sum_i\ ,{\rm etc.}\ee
Thus we have
\begin{eqnarray}\label{2.18}
\Gamma_k^\Lambda(\varphi)&=&\frac{1}{2}\varphi_i\varphi_j\Gamma_2^{ij}+
\frac{1}{4!}
\varphi_i\varphi_j\varphi_k\varphi_l\Gamma_4^{ijkl}\nonumber\\
&&+\frac{1}{6!}\varphi_i\varphi_j\varphi_k\varphi_l\varphi_m\varphi_n
\Gamma_6^{ijklmn}+...
\end{eqnarray}
Inserting this expansion into the flow equation (\ref{2.13}) and
ordering the result according to powers of $\varphi$, one obtains the
following infinite system of equations:
\begin{eqnarray}\label{2.19}
\partial_k\Gamma^{ij}_2&=&\partial_kR_k^{\Lambda,hl}[\delta_h^i
\delta_l^j-\frac{1}{2}
\Gamma_{2lm}^{-1}\Gamma_4^{mnij}\Gamma_{2nh}^{-1}]\nonumber\\
\partial_k\Gamma_4^{ijhl}&=&\partial_kR_k^{\Lambda,mn}\Bigl[3\Gamma^{-1}
_{2no}\Gamma_4
^{opij}\Gamma_{2pq}^{-1}\Gamma_4^{qrhl}\Gamma^{-1}_{2rm}
-\frac{1}{2}\Gamma_{2no}^{-1}\Gamma^{opijhl}_6\Gamma^{-1}_{2pm}\Bigr]
\nonumber\\
&.&\nonumber\\
&.&\nonumber\\
&.&
\end{eqnarray}
The r.h.s. of the flow equations for the individual $N$ point functions
$\Gamma_N$ have the following properties: In contrast to standard
$\beta$ functions each of them is exact, there are no higher order
terms beyond the ones shown explicitly. They have a simple diagrammatic
interpretation shown in fig. 1. The flow equation for $\Gamma_N$
includes always a term involving $\Gamma_{N+2}$ on its r.h.s, thus a
truncation of the series leaves no exact result. On the other hand, the
iterative solution of these differential equations generates the loop
expansion of standard perturbation theory, because the r.h.s. of each
equation involves effectively an overall factor of $\hbar$.

\section{Flow Equation for the Four Point Function}
\setcounter{equation}{0}

In this section we will consider a truncation of the flow equation of
the four point function, which corresponds to the ladder approximation
in the Bethe-Salpeter framework. First,  we will neglect the
contributions to the flow of the full inverse propagator $\Gamma_2$
induced  by the interactions and identify $\Gamma_2$ with the bare
(regularized) inverse propagator, $\Gamma_2=R_k^\Lambda$ (or
$\Gamma_2^{-1}=P_k^\Lambda)$. Second, we neglect the contribution of
$\Gamma_6$ to the flow of $\Gamma_4$, which amount to taking only
bubble-type diagrams into account.

We will slightly change the field content, namely include two distinct
complex scalar fields $\varphi_a,\varphi_b$ with identical mass $m$ for
simplicity. The expansion of the effective action reads
\be\label{3.1}
\Gamma^\Lambda_k(\varphi_a,\varphi_b)=\varphi^\dagger_{ai}R_
{k\phantom{m}j}^{\Lambda i} \varphi^j_a+\varphi^\dagger_{bi}R_
{k\phantom{m}j}^{\Lambda i} \varphi^j_b+
\varphi^\dagger_{ai}\varphi^j_a\varphi^\dagger_{bh}
\varphi^l_b\Gamma^{ih}_{4jl}+...\ee
and the flow equation for
$\Gamma_4$ becomes in the present approximation
\be\label{3.2}
\partial_k\Gamma_{4jl}^{ih}=-2\partial_kP^{\Lambda m}_
{k\phantom{m}n}
\Gamma^{nh}_{4ol}P^{\Lambda o}
_{k\phantom{m}p}\Gamma^{ip}_{4jm}.\ee
With momenta written explicitly this equation reads
\begin{eqnarray}\label{3.3}
&&\partial_k\Gamma_4(p_1,p_2,p_3,p_4)=-2\int\frac{d^4q}{(2\pi)^4}
\Gamma_4(p_1,p_2,q,-q-p_1-p_2)
\nonumber\\
&&\cdot\Gamma_4(-q,q+p_1+p_2,p_3,p_4)\cdot\partial_kP_k^\Lambda(q^2)
\cdot P^\Lambda_k ((q+p_1+p_2)^2).\end{eqnarray}

The equivalence to the Bethe-Salpeter equation can be established as
follows: Let us introduce
a two-particle propagator $P^{(2)}$
through
\be\label{3.4}
P^{(2)ij}_{\phantom{(2)ij}mn}\equiv P_{k\phantom{j}m}^{\Lambda i}
P_{k\phantom{j}
n}^{\Lambda j}.\ee
Then eq. (\ref{3.2}) can formally be written as
\be\label{3.5}
\partial_k\Gamma_4=-\Gamma_4\otimes\partial_kP^{(2)}\otimes \Gamma_4\ee
with the formal solution
\be\label{3.6}
\Gamma_4=\frac{\Gamma_4^\Lambda}{1+P^{(2)}\otimes\Gamma_4^\Lambda}\ee
where $\Gamma_4^\Lambda$ is the boundary value or bare four point
function.  In our convention, the non-amputated four point function
$\tilde\Gamma_4$, which also includes a disconnected piece, is related
to $\Gamma_4$ through
\be\label{3.7}
\tilde\Gamma_4=-P^{(2)}\otimes\Gamma_4\otimes P^{(2)}+P^{(2)}\ee
and satisfies the Bethe-Salpeter equation \cite{14}
\be\label{3.8}
\tilde\Gamma_4=P^{(2)}+P^{(2)}\otimes K\otimes\tilde\Gamma_4\ee
where $K$ denotes the interaction kernel. The formal solution of
(\ref{3.8}) reads
\be\label{3.9}
\tilde\Gamma_4=\frac{P^{(2)}}{1-K\otimes P^{(2)}}=P^{(2)}+\frac{P^{(2)}
\otimes K\otimes P^{(2)}}{1-K\otimes P^{(2)}}.\ee
Using eq. (\ref{3.7}), one finds that this solution of the B.-S.
equation coincides with the solution (\ref{3.6}) of the flow equation
after identifying $K$ with $-\Gamma^\Lambda_4$. Thus, although these
equations are quite different, they have the same physical content. We
expect that a general equivalence between the full set of flow
equations (\ref{2.19}), beyond the ladder approximation, and the full
set of Schwinger-Dyson equations can be established.

This discussion showed already an equivalence between the boundary
value of the running four point function $\Gamma_4$ at $k=\Lambda$,
denoted by $\Gamma_4^\Lambda$ in eq. (\ref{3.6}), and a general
interaction kernel $K$. In fact, $\Gamma ^\Lambda_4$ does not
necessarily have to be a bare coupling of a theory with only
$\varphi_a,\varphi_b$ fields, but could be the result of interactions
with different  fields which have already been completely integrated
out. An example is given by the Wick-Cutkosky model \cite{14}, which
starts out with an additional real massless scalar field $\phi$ beyond
the fields $\varphi_a,\varphi_b$. The original interaction involves no
bare four point coupling as in eq. (\ref{3.1}), but just trilinear
interactions of the form
\be\label{3.10}
\Gamma_{int}=g\phi(\varphi^\dagger_a\varphi_a+\varphi_b^\dagger
\varphi_b).\ee

It is straightforward  to integrate out the field $\phi$, which gives
rise to an effective (nonlocal) four point function for the field
$\varphi_a,\varphi_b$ as in (\ref{3.1}) with
\be\label{3.11}
\Gamma_4^\Lambda(p_1,p_2,p_3,p_4)=\frac{g^2}{(p_1-p_3)^2}=\frac{g^2}
{(p_2-p_4)^2}.\ee
The dynamics of the remaining fields $\varphi_a, \varphi_b$ in the
presence of the interaction (\ref{3.11}) can now either be obtained by
means of the B.S. equation or, as we propose here, by integrating the
flow equations for $\Gamma_4$ as given by eq. (\ref{3.3}) with
(\ref{3.11}) as boundary condition. It is easy to imagine more general
boundary conditions obtained, e.g., after integrating out photons,
gluons, or the like.

In order to deal with eq. (\ref{3.3}) we next observe that, since the
framework is completely Lorentz-covariant, $\Gamma_4$ should be a
function of Lorentz-invariant products of momenta only. Off-shell six
independent invariants can be formed, which we denote by
\begin{eqnarray}\label{3.12}
s=(p_1+p_2)^2=(p_3+p_4)^2,&& t=(p_2-p_3)^2=(p_2-p_4)^2\nonumber\\
v_1=p^2_1+p_2^2,\ v_2=p^2_3+p^2_4,&& w_1=p^2_1-p^2_2,\ w_2=
p^2_3-p^2_4.\end{eqnarray}
Although $\Gamma_4$ will in general be an arbitrary function of those
six variables, we will first study the case where $\Gamma_4$ can be
written as a product as follows:
\be\label{3.13}
\Gamma_4=f(v_1,w_1)\cdot D_k(s)\cdot f(v_2,w_2).\ee

If one inserts this ansatz into the flow equation (\ref{3.3}), one
finds that it is stable and thus corresponds to a ``fix point''.
Furthermore variables can be separated, and one finds that the function
$f$ can depend on $k$ only via an overall factor, which can be included
in $D_k(s)$ by definition. This latter function satisfies a flow
equation of the form
\be\label{3.14} \partial_kD_k(s)=-2D_k^2(s)\cdot
F_k(s,f)\ee with
\be\label{3.15}
F_k(s,f)=\int\frac{d^4q}{(2\pi)^4}f(q,-q-p_1-p_2)f(-q,q+p_1+p_2)
\partial_kP_k^\Lambda
(q^2)P^\Lambda_k((q+p_1+p_2)^2)\ee
The general solution of the flow equation (\ref{3.14}) is
easily written down in terms of an arbitrary
boundary condition $D_{k'}(s)$ at $k=k'$:
\be\label{3.16}
D_k(s)=\frac{D_{k'}(s)}{1-2D_{k'}(s)\int^{{k'}^2}_{k^2}d\hat
k^2F_{\hat k}(s,f)}\ee
On the one hand, this shows that any problem where $\Gamma_4^\Lambda$
starts out to be of the form (\ref{3.13}) is analytically solvable. On
the other hand, \underline{iff} the theory contains a bound state in
the channel $(\varphi_a,\varphi_b) \to(\varphi_a,\varphi_b)$, which
manifests itself as a pole in $s$ of $\Gamma_4$, we actually expect
that $\Gamma_4$ approaches the form (\ref{3.13}) for $k\to0$ and $s$ in
the vincinity of the pole. Then the function $f$ corresponds to the
amputated Bethe-Salpeter wave function of the bound state, whereas
$D_k(s)$ for $k\to0$ corresponds to the bound state propagator and
contains the information on the position of the
pole. In fact we will see that the factorized form (\ref{3.13}) of
$\Gamma_4$ is an infrared-attractive ``fix point'' of the flow
equations, which will be of great help in extracting informations on
possible bound states analytically.

In order to proceed towards concrete calculations we have to make a
choice on the form of the regularization of the propagator
$P_k^\Lambda$ or on the function $h_k(q^2)$ of eqs. (\ref{2.1}) and
(\ref{2.2}). The following choice turns out to be particularly useful:
\be\label{3.17} h_k(q^2)=e^{-\frac{q^2+m^2}{k^2}}\ee
or
\be\label{3.18}
P_k^\Lambda(q^2)=\frac{e^{-\frac{q^2+m^2}{\Lambda^2}}-e^
{-\frac{q^2+m^2}{k^2}}}{q^2+m^2}=
\int^{1/k^2}_{1/\Lambda^2}d\alpha e^{-\alpha(q^2+m^2)}.\ee
This regularized propagator has the following properties:
%For $k^2\llm^2$, it remains finite also for $q^2\ll k^2$.
For general nonvanishing
$k^2$, there is no pole at $q^2=-m^2$, instead it behaves like $1/k^2$,
at $q^2=-m^2$, for $k^2\to 0$. Beyond the ``pole'', for $q^2<-m^2$, it
increases exponentially for $k^2\to 0$. The absence of singularities at
finite $k^2$ is also clear from the boundedness of the Feynman
parameter integral $d\alpha$.

Similarly we have to make a choice on the parametrization of the
dependence of $\Gamma_4$ on the six Lorentz invariants shown in
(\ref{3.12}). First we make a technical simplification, namely we
neglect the dependence of $\Gamma_4$ on the variables $w_1,w_2$ of
(\ref{3.12}). In the Wick-Cutkosky model $\Gamma_4$ does not depend on
$w_1,w_2$ neither at its starting point at $k\to\Lambda$, given by eq.
(\ref{3.11}) in the form $g^2/t$, nor for $k\to0$ in the weak coupling
limit. In this limit the wave function $f$ is known to depend on $v$
only.

Now it turns out to be very convenient to switch from
$\Gamma_4(s,t,v_1,v_2)$ to its Laplace transform with respect to the
three variables $t,v_1$ and $v_2$:
\be\label{3.19}
\Gamma_4(s,t,v_1,v_2)=\int^\infty_0dl_0dl_1dl_2C_k(s,l_0,l_1,l_2)
e^{-l_0t-l_1v_1-l_2v_2}\ee
After inserting this Laplace transformation together with the
parametrization (\ref{3.18}) of the propagator into the flow equation
(\ref{3.3}), one finds that the $d^4q$ integration can easily be
performed, and one is left with the following flow  equation for $C_k$:
\begin{eqnarray}\label{3.20}
&&\partial_kC_k(s,l_0,l_1,l_2)=\frac{1}{8\pi^2k^2}\int^1_{k^2/\Lambda^2}
d\beta\int^\infty_0dn_0 dn_1dn_2dm_0dm_1dm_2\nonumber\\
&&\cdot\frac{1}{B^2}\cdot
e^{-\frac{1}{k^2}(m^2(1+\beta)+s\frac{bb'}{B})}
\delta(l_0-\frac{k^2n_0m_0}{B})\delta(l_1-\frac{n_0(b+b')}{2B}-n_1)
\nonumber\\ &&\cdot\delta(l_2-\frac{m_0(b+b')}{2B}-m_2)\cdot
C_k(s,n_0,n_1,n_2)\cdot C_k(s,m_0,m_1,m_2) \end{eqnarray}
with
\be\label{3.21}
b=1+k^2(n_2+m_1),\ b'=\beta+k^2(n_2+m_1),\ B=b+b'+k^2(n_0+m_0).\ee
Of course the previously discussed simplification in the case of
factorization remain valid: If $\Gamma_4$ factorized as in eq.
(\ref{3.13}), $C_k$ factorizes as
\be\label{3.22}
C_k(s,l_0,l_1,l_2)=\delta(l_0)\tilde f(l_1)D_k(s)\tilde f(l_2)\ee
where $\tilde f(l)$ is the Laplace transform of the wave function
$f(v)$. The solution (\ref{3.16}) for $D_k(s)$ remains the same with
$F_k(s,f)$ now given by
\be\label{3.23}
F_k(s,f)=\frac{1}{16\pi^2k^2}\int^1_{k^2/\Lambda^2}d\beta\int^
\infty_0dl_1dl_2\frac{
\tilde f(l_1)\tilde f(l_2)}{B^2}e^{-\frac{1}{k^2}(m^2(1+\beta)+s
\frac{bb'}{B})}\ee
and $b=1+k^2(l_1+l_2),\ b'=\beta+k^2(l_1+l_2),\ B=b+b'$.
The right-hand side of eq. (\ref{3.20}) shows already an expected
analytic property of the four point function: If we continue the
variable $s$ into the Minkowskian regime $(s<0)$, we expect a cut at
$s=-4m^2$. Indeed, in the limit $k^2\to0$ and for $\beta\to1$ the
exponent on the r.h.s. of eq. (\ref{3.20}) becomes $-(2m^2+s/2)/k^2$
and thus explodes for $s<-4m^2$.

In the case of general interactions, i.e. general boundary conditions
for $\Gamma_4$ or $C_k$ at $k=\Lambda$, the flow equations either in
the form (\ref{3.3}) or in the form (\ref{3.20})   cannot be solved
analytically. Now we propose to solve eq. (\ref{3.20}) numerically
after discretization of the variables $l_0,l_1$ and $l_2$.
$C_k(s,l_0,l_1, l_2)$ becomes a function $C_k(s)_{i_0i_1i_2}$, which
lives on a three-dimensional lattice, where the integers $i_0,i_1,i_2$
denote the lattice points. After discretization of the variable $k^2$
eq. (\ref{3.20}) provides an algorithm on how to update the lattice in
each step from $C_k(s)$ to $C_{k-\Delta k}(s)$.

It is sensible to start this algorithm with values of $k$ large
compared with the scale $m$ of the problem; choosing the starting point
$\Lambda$ to be $\Lambda^2\sim 10 m^2$ turns out to be sufficient for
our purposes. At the beginning one has to fix $C_\Lambda$, the Laplace
transform of $\Gamma_4^\Lambda$, and the variable $s$.

In the presence of a bound state with mass $M$ we expect the following
behaviour of $C_k$ in analogy to the behavior of the regularized
propagator  $P_k^\Lambda$ of eq. (\ref{3.18}):  For $s>-M^2\ C_k$
remains finite for $k\to 0$, whereas for $s\lta -M^2\ C_k$ approaches a
factorized form as in eq. (\ref{3.22}), and the factor $D_k(s)$
diverges for $s<-M^2$ and $k\to 0$. A divergence of $D_k$ for $k \to 0$
thus indicates, that the variable $s$ satisfies $s<-M^2$. This
behaviour is indeed the result  of our numerical investigations.

Since the factorized form (\ref{3.22}) turns out to be infrared stable,
eq. (\ref{3.16}) can help us to decide whether,  at a given value of
$s,\ C_{k\to0}$ diverges or not: If we observe that for a certain value
of $k=k'$, $C_{k'}$ assumes the form (\ref{3.22}), we can read off the
function $\tilde f(l)$ and the value of $D_{k'}(s)$. Then we can
compute the function $F_k(s,f)$ of eq. (\ref{3.23}), insert it into
eq. (\ref{3.16}), and we see immediately that $D_{k\to0}^{(s)}$
diverges if
\be\label{3.24} 2D_{k'}(s)\int^{{k'}^2}_0d\hat k^2F_{\hat
k}(s,f)>1.\ee
There is thus no need to integrate numerically down to $k=0$, but the
following procedure turns out to be sufficient: We integrate
numerically the flow equation for $C_k$, at a given value of $s$, down
to $k^2={k'}^2\simeq m^2/10$. Then we check whether $C_k$ has assumed
the factorized form (\ref{3.22}). If not, this remains so even for
smaller values of $k^2$ (as we have tested numerically) and we
conclude, that the corresponding value  of $s$ satisfies $s>-M^2$,
where $M$ is the mass of the lowest lying bound state. Thus we decrease
$s$ and start again, until we get to a value of $s$ where $C_{k'}(s)$
factorizes. Then we read off $D_{k'}(s)$ and $f(l)$ and check with the
help of eq. (\ref{3.24}) whether  $D_{k\to0}(s)$ diverges, i.e. whether
$s$ satisfies $s<M^2$. By repeating this procedure for different values
of $s$, we can pin down $M^2$ as accurately as we like, and of course
we have also obtained the Laplace-transformed wave function $\tilde
f(l)$.

The general picture of the use of flow equations is the following: For
$k^2\gg m^2$,
 the r.h.s. of the flow equations is small except for Green functions
which correspond to marginal or relevant operators, which are thus the
only ones to vary significantly (but which are not present in the
present approximation).
In this regime,  at least for small couplings, the result of the
integration of the flow equaitons could also be obtained by standard
perturbative methods, eventually improved by the use of any
mass-independent standard  renormalization group. For $k^2\sim m^2$,
however, dynamical effects like the formation of bound states take
place, which generally require the use of numerical methods. For
$k^2\ll m^2$, the r.h.sides of the flow equations are again small
(damped exponentially) except for external momenta, where Green
functions develop singularities like poles or cuts. This is a feature
both of the bare propagator (\ref{3.18}) and of the four point function
discussed here. The development of singularities, however, has simple
universal properties like the factorized behaviour (\ref{3.13}) of the
four point function, which allow it to be treated analytically again.

The general procedure of combined numerical and analytic methods for
the search for bound states as described above will be applied to some
concrete models in the next section.

\section{Results for Bound States}
\setcounter{equation}{0}
In this chapter we will study the bound  state problem in the case of
two complex scalar fields using the methods derived in section 3. We
integrate the flow equations for the four point function $\Gamma_4$ in
an approximation, where the scalar propagator remains the free one, and
where the contribution of the six point function to the r.h.s. of the
flow equation for $\Gamma_4$ is neglected. After restricting the
momentum dependence of $\Gamma_4$ to the four invariants $s,t,v_1$ and
$v_2$ of (\ref{3.12}) and after the Laplace transformation with respect
to $t, v_1$ and $v_2$, the flow  equation has the form (\ref{3.20}).

The information on the underlying model, i.e. the bare interaction
$\Gamma_4^\Lambda$ or the kernel of the corresponding Bethe-Salpeter
equation, is entirely and only  specified by the boundary condition at
$k=\Lambda$. In the case of the Wick-Cutkosky model, which we study
first, this boundary condition is given by $\Gamma^\Lambda_4 =g^2/t$
(see eq. (\ref{3.11}) or, in terms of the Laplace transform $C_k$ of
(\ref{3.19}), by
\be\label{4.1}
C_\Lambda(s,l_0,l_1,l_2)=g^2\delta(l_1)\delta(l_2).\ee
As described in
the previous chapter, we discretisize the variables $l_0,l_1$ and
$l_2$. We  used lattice sizes up to 20 in all three directions. We
integrate the flow equation (\ref{3.20}) numerically from
$k^2=\Lambda^2=10 m^2$ down to $k^2={k'}^2=m^2/10$ at fixed c.m.
energy  $s$ with $-4m^2<s<0$. At $k^2={k'}^2$ we check, whether
$C_{k'}$ has  at least approximately assumed the factorized form
(\ref{3.22}). If this is the  case, we extract the function $\tilde
f(l)$ and the value of $D_{k'}(s)$. Then we compute the l.h.s. side of
(\ref{3.24}), still at fixed $s$, in order to check whether $s$
satisfies $s>-M^2$ or $s<-M^2$, where $M$ is the mass of the
lowest-lying bound state.

Indeed this method works very well in the present model. If $s$ is
chosen close to or below $-M^2$, the numerical deviations of $C_k$,
from the factorized form (\ref{3.22}), are  less than one percent. For
such values of $s$, $C_k$ is seen to increase rapidly towards small
values of $k$. For larger values of $s$, on the other hand, $C_k$ is
seen to remain finite for $k\to 0$. If we would take
$C_{k'}(s,0,0,0)>10^5$ as a criterium for $s<-M^2$, this would actually
match our criterium for $s<-M^2$ based on eq. (\ref{3.24}) already
within a few percent of  the value of s.

Within the Wick-Cutkosky model it is convenient to study the dependence
of the mass $M$ of the lowest lying bound state on a coupling $\lambda$
defined by
\be\label{4.2}
\lambda=g^2/16\pi m^2.\ee
Analytic results are known in the weak coupling limit $\lambda\to0$,
where
\be\label{4.3}
M^2\simeq(4-\lambda^2)m^2,\ee
and in the case of a vanishing bound state mass \cite{14}:
\be\label{4.4}
M=0\quad{\rm for}\quad \lambda=2\pi.\ee
A result of a numerical solution of the Bethe-Salpeter equation can
be found in \cite{15}:
\be\label{4.5}
M=(2-.082)m\quad {\rm for}\quad \lambda =1.\ee
In fig. 2 we plot our results for $M^2$ in units of $m^2$ versus
$\lambda$. We indicate, for $\lambda <.3$, eq. (\ref{4.3}) as a short
line, and the two results (\ref{4.4}) and (\ref{4.5}) as crosses. We
show our results as error bars, which were obtained on a $20^3$
lattice. The
error bars are due to varying the prescriptions on how to discretisize
the variables $l_0,l_1$ and $l_2$, and due to changing the physical
size of the  lattice: The ranges of $l_0$, $l_1$  and $l_2$ are given
by  $0\leq l_0\leq l_{0max}$ and $0\leq l_1,l_2\leq l_{max}$, and we
varied $l_{0max}$ between 10 and 40, and $l_{max}$ between 6 and 10 (in
units of $m^{-2}$). (Larger values of $l_{max}$ are irrelevant, because
$C_k(s,l_0,l_1,l_2)$ decreases exponentially with $l_1$  and $l_2$, see
below.)

In the case of large values of the coupling $\lambda$ it should be
remembered that we neglected the dependence of $\Gamma_4$ on the
momentum variables $w_1,w_2$ of eq. (\ref{3.12}). This is known to be a
good approximation, for $\Gamma_4$ with $k\ll\Lambda$, only in the
nonrelativistic or weak coupling limit. Therefore the deviations of our
results from the known behaviour (\ref{4.4}) for $\lambda \to 2\pi$
(strong coupling) are understandable.

In the extreme weak coupling limit $\lambda\to0$, on the other hand,
the physical size of a bound state, in ordinary space, is known to
increase. In terms of our  ``Laplace space'', spanned by the variables
$l_i$, large distance also correspond to large values of $l_i$. Since
this ``Laplace space'' necessarily has a finite size, it is  again
understandable that this method cannot describe phenomena accurately,
where very large distances play an important role, as in the case of
very weakly coupled  bound states.

For moderate values of $\lambda$ we find  a large region, however,
where our  results seem to be free of those systematic errors and
match the known ones suprising well. The dependence
of our results on the lattice size might be of some interest. Therefore
we plot, for $\lambda=1$, our results for $M^2$ versus the lattice size
in fig. 3. (The error bars are of the same origin as discussed above.)
We see the steady convergence, with increasing lattice size, towards
the known result, which makes us believe that the method has no
systematic limitations beyond the ones mentioned before.

The procedure required already to extract the function $\tilde f(l)$
out of $C_{k'}$, which satisfies (\ref{3.22}) for $s$ near $-M^2$. Up
to a normalization this function is the Laplace transform of the
amputated Bethe-Salpeter wave function. In the weak coupling limit of
the Wick-Cutkosky model this amputated wave function  is known to be
\cite{14}
\be\label{4.6}
f(p^2_1,p^2_2)=1/(p^2_1+p^2_2+2m^2)=1/(v_1+2m^2)\ee
with $v_1$ as in (\ref{3.12}). Accordingly $\tilde f(l)$ should read
\be\label{4.7}
\tilde f(l)=\exp(-2m^2l).\ee
In fig. 4 we plot our result for $\log[\tilde f(l)]$, at $\lambda=1$,
versus $l$ in units of $m^{-2}$ for a lattice size of 10. We normalized
our result with respect to the known one at $l=1$, and we see very
nicely the agreement with the known exponential decrease, until finite
lattice size effects start to play a role.

It should be noted that, in order to obtain all the results of figs. 2
to 4 from the numerical integration of the flow equation (\ref{3.20}),
a computer time on a work station of only a few minutes was needed.
Furthermore no use was made of the $O(4)$ symmetry of the Wick-Cutkosky
model \cite{14}, which was exploited in the numerical solution of the
Bethe-Salpeter equation in \cite{15}. Thus the method can
straightforwardly be applied to the relativistic bound state problem
with arbitrary interactions.

Of particular interest are interaction kernels, which are motivated by
QCD and which correspond to a confining potential in the
nonrelativistic limit. Let us first discuss potentials $V(r)$, which
behave like $r^\alpha$ as a function of the distance $r$ in ordinary
three-dimensional space.  The corresponding Lorentz-invariant
momentum-dependent interaction kernel is actually not unique; it
becomes unique, however, if one assumes that it depends only on the
momentum transfer $q^2=t$. We adopt the notation to our procedure,
where the interaction kernel coincides with the boundary value
$\Gamma^\Lambda_4$ of the four point function; then the relation
between $V(r)$ and $\Gamma_4^\Lambda(t)$ reads
\be\label{4.8}
\Gamma_4^\Lambda(t)=\gamma 2^{3+\alpha}\pi^{3/2}t^{-3-\alpha}\frac
{\Gamma(\frac{3+\alpha}{2})}{\Gamma(-\frac{\alpha}{2})}\quad {\rm for}
\quad V(r)=\gamma r^\alpha.\ee
In our formalism we actually need the Laplace transform $C_\Lambda$ in
analogy to (\ref{4.1}); its corresponding form is given by
$C_\Lambda(s,l_0,l_1,l_2)= C_\Lambda(l_0)\delta(l_1)\delta(l_2)$ with
\be\label{4.9}
C_\Lambda(l_0)=\gamma \frac{2^{3+\alpha}\pi^{3/2}}{\Gamma(-\frac
{\alpha}{2})}l_0^{\frac{1+\alpha}{2}}.\ee
The possible singularities in $\alpha$ in eqs. (\ref{4.8}), (\ref{4.9})
originate from analytic continuation  in $\alpha$, which is required if
the functions $V(r)$ or $C_\Lambda(l_0)$ are not  well behaved at large
distances $r\to\infty$ or $l_0\to\infty$. Within our procedure,
however, $C_\Lambda(l_0)$ will only be nonvanishing for a finite range
of $l_0$ due to the finite size of the lattice. From the general
relation between $V(r)$ and $C_\Lambda (l_0)$,
\be\label{4.10}
V(r)=\frac{1}{4\pi^{3/2}r}\int^\infty_0\frac{dx}{\sqrt x}C_\Lambda(l_0)
e^{-x}\quad
{\rm with}\quad l_0=\frac{r^2}{4x},\ee
one finds the following: If $C_\Lambda(l_0)$ vanishes for $l_0\geq a,\
V(r)$ decays exponentially for $r^2\gg a$. On the other hand, for
$r^2\ll a, \ V(r)$ and $C_\Lambda(l_0)$ are related as given by eqs.
(\ref{4.8}), (\ref{4.9}) up to a constant in $V(r)$ depending on $a$.
Such a  constant is actually quite welcome; mesonic spectra in QCD, in
the nonrelativistic limit, are typically derived from potentials of the
form
\be\label{4.11} V(r)=\lambda r-\frac{4}{3}\alpha_{QCD}r^{-1}-c\ee
with a constant $c$ of $O$(1 GeV) \cite{16}.

We applied our method to relativistic interactions, which are motivated
by the potential (\ref{4.11}). Of course we are dealing with scalars
instead of fermionic quarks, but scalars are known to reproduce the
spectrum of fermionic bound states in the nonrelativistic limit (where
spin effects become negligible), provided the Coulomb interaction
induced by vector boson exchange is multiplied by a factor 4.

Thus we integrated the flow equations (\ref{3.2}) with a boundary
condition $C_\Lambda(l_0)$ of the form
\be\label{4.12}
C_\Lambda(l_0)=-\theta(\frac{\pi c^2}{4\lambda^2}-l_0)(8\pi\lambda l_0+
16\pi\cdot\frac{4}{3}\alpha_{QCD}),\ee
which reproduces (\ref{4.11}) for $r\ll c/\lambda$ according to
(\ref{4.10}). For the string tension $\lambda$ we choose $.25$
${\GeV}^2$, $.3$ for $\alpha_{QCD}$ and  1 {\GeV} for the constant
$c$. In table 1 we give our results for the bound state mass $M$ for
different values of the scalar mass $m$ in the range 1.5 {\GeV} to  5.5
{\GeV}. Again the results agree with general expectations: Small, but
increasing binding energy towards larger scalar masses, and vanishing
binding energy for $m\lta 1 {\GeV}$ \cite{16}. (Our present method does
not allow us to find bound states above threshold or with $M>2m$, since
for $s<-4m^2$ the behaviour of the four point function is dominated by
the  two-particle cut, which reveals itself as a singularity  for
$k\to0$.) The increasing errors towards larger scalar masses and
binding energies likely indicate that here the neglect of the
dependence of $\Gamma_4$ on the momentum variables $w_1,w_2$ is not a
good approximation. Nevertheless the results show certainly the
feasibility of the  method in the case of general interactions.

\section{Conclusions}

In this paper we presented a derivation of the exact flow equation
(\ref{2.13}), which  allows to compute effective low energy actions in
terms of arbitrary high energy actions. It corresponds to an infinite
set of flow equations for one particle irreducible Green functions,
where each equation is again exact. Solutions to a finite subset of
these equations are already nonperturbative in $\hbar$, and correspond
to summations of certain sets of Feynman diagrams. They contain much
more information, however, than the finite number of running coupling
constants in standard renormalization theory.

Whereas the practical use of flow equations, often in different
formulations, has been demonstrated before in context of expansions of
effective actions in powers of momenta \cite{10}-\cite{13}, we
concentrated here on the bound state problem. Our tool was a truncated
flow equation for the four point function, corresponding to the
summation of ``ladder type'' diagrams, with the full dependence on the
momenta left intact.

In principle, the effective low energy action and hence the low energy
four point function contains all informations available by summing
perturbation theory. We presented some technical tools, which allow to
make practical use of this principle. These  tools involved the
parametrization of the momentum dependence of Green function in terms
of its Laplace transform with respect to the Lorentz-invariant products
of momenta (except for the c.m. energy $s$). This procedure is
obviously  manifest Lorentz-invariant, and simplifies the r.h.s. of
the  flow equations such that they become accessible to a numerical
treatment.

One has not to rely completely on numerical methods, however: In the
presence of nontrivial analytic structures, such as bound state poles,
the form of the Green functions simplifies at small scales such that
the flow equations can be solved analytically in this regime. The
convergence of the Green functions towards these simple structures (as
the factorized form (\ref{3.13})) in the infrared represents a new form
of universality.

We demonstrated the practical feasability of the method in the framework
of the Wick-Cutkosky model, which represents a non-trivial bound-state
problem with some known results derived via the Bethe-Salpeter
equation. We checked the possibility to continue the c.m. energy $s$
towards negative values (into the Minkowski regime), and the
possibility to obtain both the mass of the lowest-lying bound state as
well  as the corresponding wave function with the help of the proposed
combination of  numerical and analytic methods.

The general applicability of these methods allowed us to investigate
very different interactions such as QCD-motivated confining kernels.
The relation between the Laplace transform of such kernels and
confining nonrelativistic potentials including a constant term is
actually an interesting subject for its own, we just briefly presented
some general formulas on this relation. Again our results look quite
promising, but this field deserves further studies.

Clearly our approach can be improved and extended in a straightforward
manner: The dependence of the four point function on the momentum
variables $w_1,w_2$ can be restored, and fermions can be included
\cite{17}.
Furthermore the flow equation for the two point function or the exact
propagator can be taken into account, and it is possible to go beyond
the ladder approximation by including the six or even higher point
functions. Also the consideration of gauge interactions will impose no
intractable problems \cite{18}: Of course the momentum cutoff has to be
covariantized, which is straightforward since the UV cutoff $\Lambda$
remains fixed throughout the whole procedure. It will turn out,
however, that an additional one-loop counterterm is required \cite{18};
this field is the subject of present investigations.

In conclusion, besides a derivation and discussion of the flow
equations we have shown that the bound-state problem is another field
of practical applicability of this approach. It seems to be general and
flexible enough to allow for studies of a wide range of phenomena in
quantum field theory.

\section*{Acknowledgement}
It is a pleassure to thank D. Gromes and C. Wetterich for
stimulating discussions.

\vspace{3cm}
\section*{Figure Captions}
\begin{description}
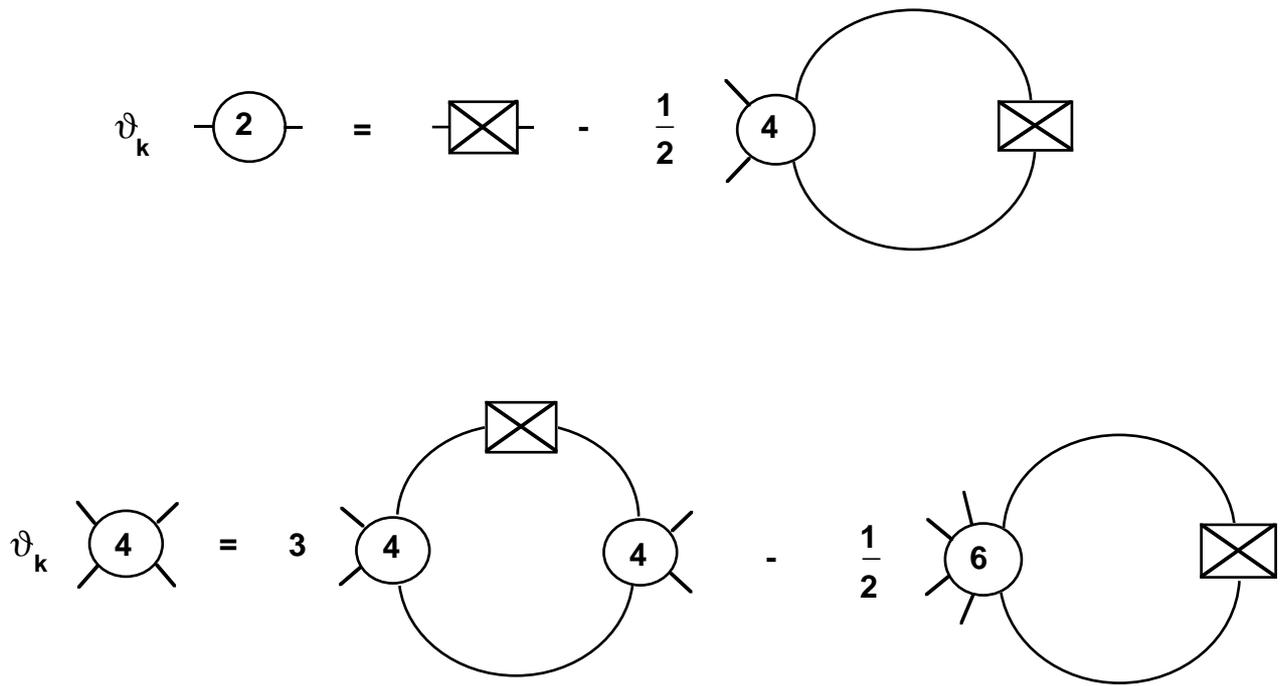
\item{Fig. 1:}
Diagrammatic representation of the flow equations for the two and four
point functions as in eq. (\ref{2.19}). The lines denote the full
propagators  $\Gamma_2^{-1}$, encircled numbers the corresponding
amputated $N$ point  function, and the cross in a box an insertion of
$\partial_kR_k^\Lambda$. \bigskip
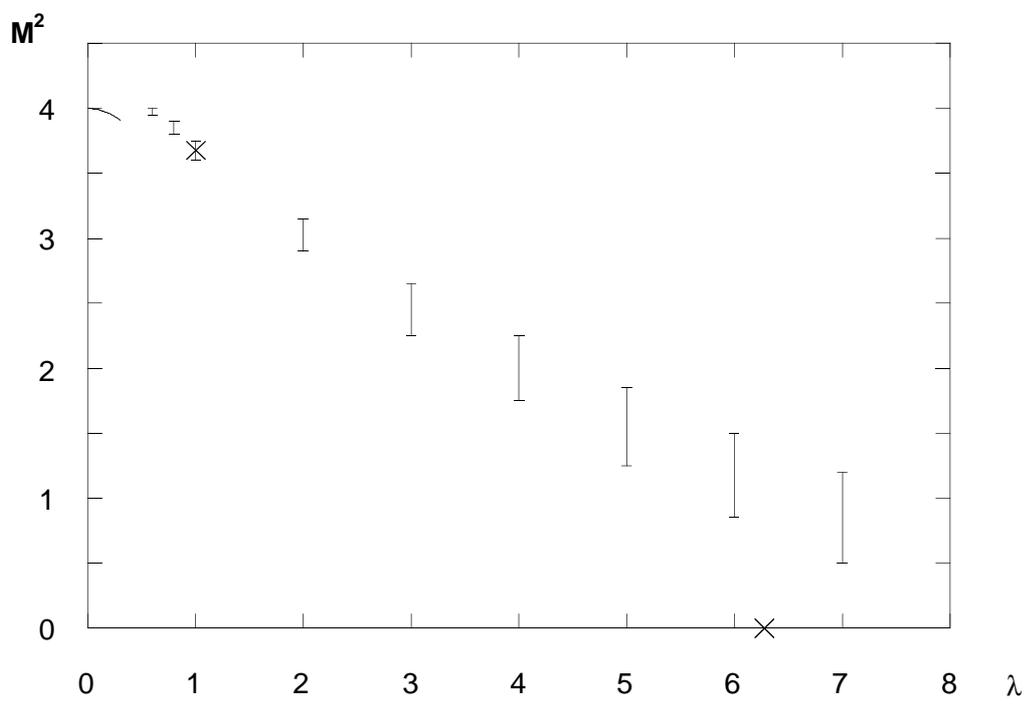
\item{Fig. 2:}
Plot of $M^2$ of the lowest-lying bound state in the Wick-Cutkosky
model versus the coupling $\lambda$. The short line denotes the known
analytic result in the weak coupling limit (see eq.  (\ref{4.3})), and
the crosses known results for  $\lambda=1$ (eq. (\ref {4.5})) and
$\lambda=2\pi$ (eq. (\ref{4.4})). Our results are shown as error bars.
\bigskip
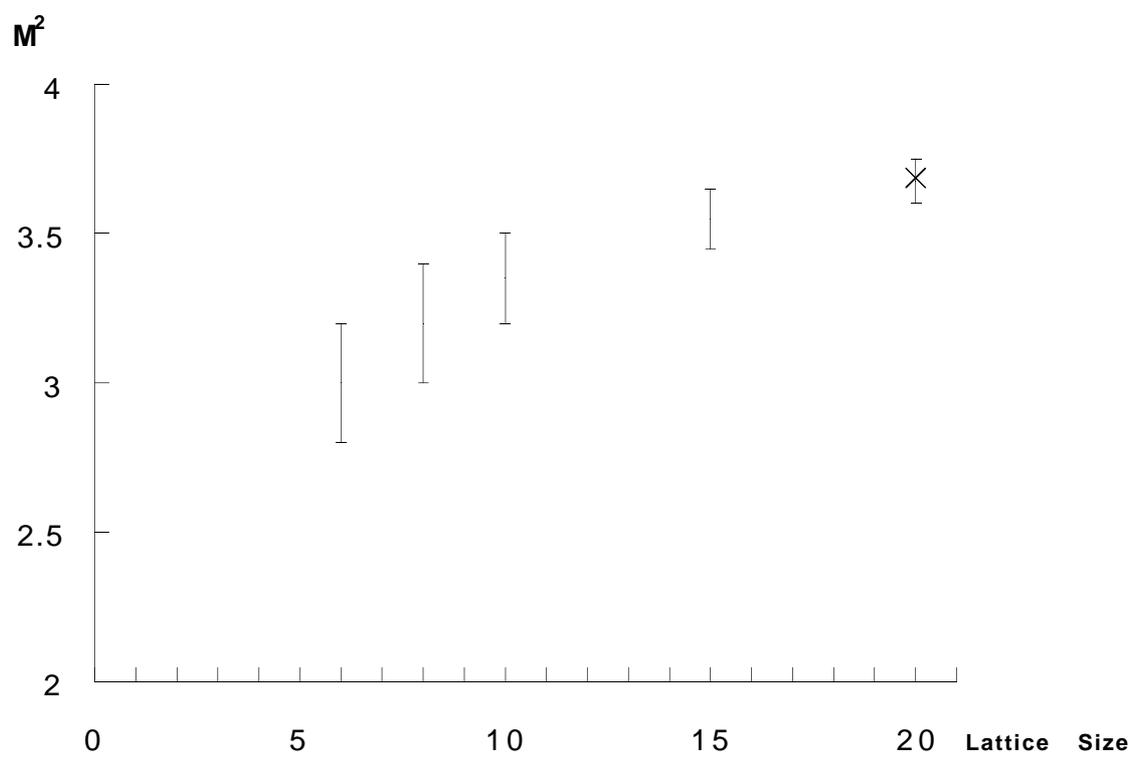
\item{Fig. 3:}
Plot of $M^2$, in the same model and for $\lambda=1$, versus the
lattice size. The known result (eq. (\ref{4.5})) is indicated as a
cross.
\bigskip
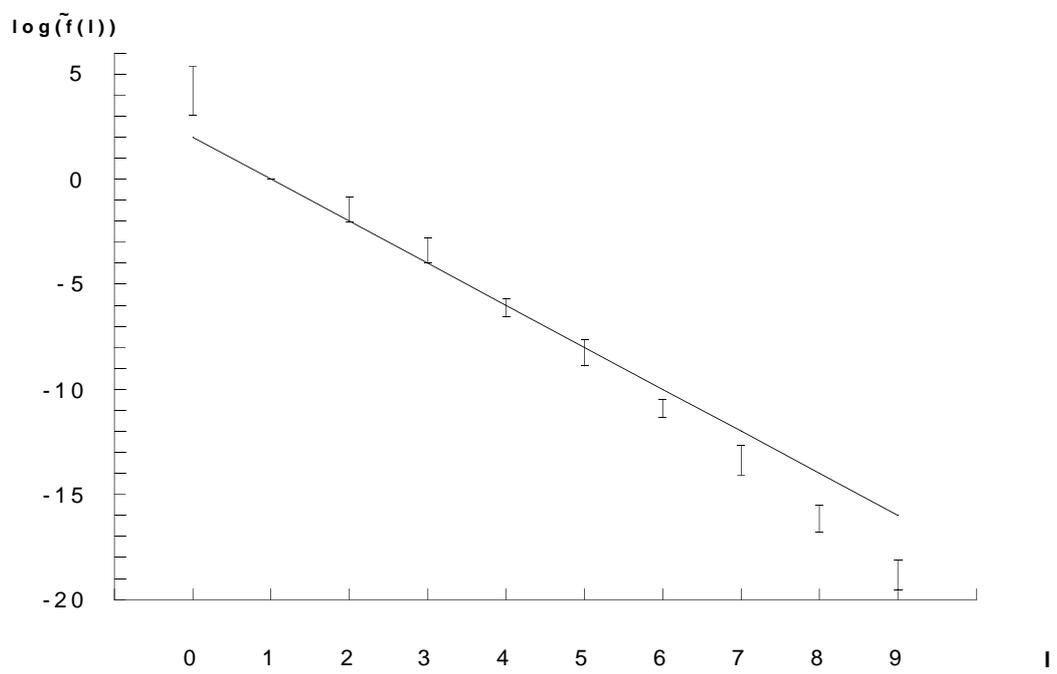
\item{Fig. 4:}
Plot of the logarithm of the Laplace transform of the wave function
$\tilde f(l)$ of the Wick-Cutkosky model in the weak coupling limit.
The known result (eq. (\ref{4.7})) is indicated as  a straight line,
and our results are normalized to the known one at $l=1$.
\end{description}

\newpage
% No number on this page
\begin{titlepage}

\begin{center}
{\bf Table 1}\\
\bigskip
\begin{tabular}{|c|c|}
\hline
$m$[GeV]& $M[\GeV]$\\
\hline
1.5&2.99--3\\
\hline
2.5&4.92--4.99\\
\hline
3.5&6.81--6.96\\
\hline
4.5&8.63--8.94\\
\hline
5.5&10.11--10.83\\
\hline
\end{tabular}\\[1cm]
\end{center}
\bigskip

Result for the bound state mass $M$ for an interaction corresponding to
a  nonrelativistic potential (\ref{4.11}) and parameters as indicated
below eq. (\ref{4.12}).  $m$ denotes the different masses of the scalar
constituents.

\end{titlepage}

\newpage
% No number on this page
\begin{titlepage}

\end{titlepage}

\end{document}